\documentclass[epj]{svjour}

\usepackage{amsmath,amsfonts}
\usepackage{epsfig}    
\usepackage{graphicx}

\def\a{\alpha}
\def\b{\beta}
\def\g{\gamma}

\def\ve{\varepsilon}
\def\sq2{Y_{0}}

\def\ve{\varepsilon}
\def\wa{\widetilde{\alpha}}

\begin{document}

\title{Magnetoelectric Cr$_2$O$_3$ and relativity
  theory
}

\author{Friedrich W. Hehl\inst{1,2,}\thanks{email: {\tt
      hehl@thp.uni-koeln.de} (corresponding author)}\and Yuri
  N. Obukhov\inst{3,}\thanks{email: {\tt yo@thp.uni-koeln.de}}\and
  Jean-Pierre Rivera\inst{4,}\thanks{email: {\tt
      Jean-Pierre.Rivera@unige.ch}}\and Hans
  Schmid\inst{4,}\thanks{email: {\tt Hans.Schmid@unige.ch}}}

\institute{Institute for Theoretical Physics, University of Cologne,
  50923 K\"oln, Germany\and Department of Physics and Astronomy,
  University of Missouri, Columbia, MO 65211, USA\and
  Department of Theoretical Physics, Moscow State University, 117234
  Moscow, Russia\and Department of Inorganic, Analytical and Applied
  Chemistry, University of Geneva, Sciences II, 1211 Geneva 4,
  Switzerland}

\date{06 March 2009, {\it file SantaBarbaraP11.tex}}

\abstract{Relativity theory is useful for understanding the
  phenomenology of the magnetoelectric effect of the antiferromagnet
  chromium sesquioxide Cr$_2$O$_3$ in two respects: (i) One gets a
  clear idea about the physical dimensions of the electromagnetic
  quantities involved, in particular about the dimensions of the
  magnetoelectric moduli that we suggest to tabulate in future as
  dimensionless relative quantities; (ii) one can recognize and
  extract a temperature dependent, 4-dimensional pseudoscalar from the
  data of magnetoelectric experiments with Cr$_2$O$_3$. This
  pseudoscalar piece of Cr$_2$O$_3$ is odd under time reflections and
  parity transformations and is structurally related (``isomorphic")
  to the gyrator of electric network theory, the axion of particle
  physics, and the perfect electromagnetic conductor of electrical
  engineering.}



\maketitle

\section{Introduction} 

Magnetism is caused by moving electric charge, and the {\it motion} of
an object is closely related to relativity theory; at least if the
charge moves with a speed $v$ that cannot be neglected as compared to
the speed of light $c$. Accordingly, in the theory of
magnetoelectricity that deals with materials where a magnetisation $M$
is induced by an electric field $E$ or a polarization $P$ by a
magnetic field $B$ (or a magnetic excitation $H$), magnetic and
electric fields are closely intertwined, and it is wise to put the
electrodynamical theory in a relativistic form right away; and this
all the more since Maxwell's theory is known to be a relativistic
theory covariant under all coordinate transformations (see Post
\cite{Post1962}).

\section{Physical dimensions in magnetoelectrics}

\subsection{Permittivity and permeability}

Before one formulates a physical theory, in the case under
consideration electrodynamics of material media, one has to make a
basic decision: Do we want to formulate the laws in terms of
``quantity equations'' or in terms of ``numerical equations''? A
quantity $A=\{A\}[A]$ is always a ``numerical value'' $\{A\}$ times a
``dimension'' $[A]$; and quantity equations interrelate these
quantities. By their very definition quantity equations are valid in
any system of units. Thus, if one puts the Maxwell equations in such a
form, they are valid in SI (the international system of units), in the
Gauss system of units, etc. By contrast, in the book of Jackson
\cite{Jackson}, for instance, on the top of each second page, it is
indicated for which system of units (SI or Gaussian) the equations
displayed are valid. Clearly, this distracts from the universal nature
of the Maxwell equations---and we will not follow that
convention. Also the elementary particle physicist prefer to formulate
the laws in such a way that they are only valid in a certain system of
units, with $\hbar=1$, $c=1$, etc. 

In order to recognize the archetypal structure of the Max\-well
equations, one should write them, as already Max\-well has done
himself---who also discusses physical dimensions in quite some
detail---in terms of quantity equations, see Post \cite{Post1962} or
\cite{Birkbook}. We will follow Post in this context.

In the past these matters of dimensions and units caused heated
debates. One can get a good idea of this from the article of Brown
\cite{WFBrown}. He gives a good and complete description of how to
change from one system of units to another one. However, in contrast
to Post, he argues that choice of a dimension is more or less a
convention. We disagree with this position. In our opinion the
dimension of a quantity encodes the operational procedure of how to
measure this quantity. 

The dimension of a quantity is an intrinsic and defining property of a
quantity. We cannot, by suitable units, kill the dimension of a
physical quantity. A dimension is something invariant.  The electric
excitation $\vec{D}$, for instance, can be measured in vacuum with the
help of the Maxwellian double plates implying the dimension
$[\vec{D}]={{\rm charge}}/{{\rm area}}={q}/{\ell^2}$, where $q$
denotes the dimension of charge and $\ell$ the dimension of
length. Similarly the electric field strength $[\vec{E}]={\rm
  force}/{\rm charge}=(w/\ell)/q=(UIt/\ell)/(It)\,$ $={U}/{\ell}$,
with $w$, $U$, $I$, and $t$ as dimensions of work (energy), voltage,
current, and time, respectively. Clearly then, quite independently of
the system of units chosen, the permittivity defined by the ratio of
the electric excitation and the electric field $\varepsilon =
\vec{D}/\vec{E}$ has the intrinsic dimension $[\varepsilon] =
\left[{\vec{D}}]/[{\vec{E}}\right]=({q}/{\ell^2})\,
{\ell}/{U}=({It}/{\ell^2})\,{\ell}/{U}={1}/(Rv)$, where $R$ and $v$
are the dimensions of resistance and velocity, respectively. In
particular, also in {\it vacuum} $[\varepsilon_0] = Y/v$, with $Y=
1/R$ as the dimension of admittance.  Note that we talk about
dimensional analysis, not about a system of units. Our discussion
looks like one relating to SI; however, this is not the case---rather
SI also uses quantity equations and hence some equations in SI
resemble those we discuss here. In the Gaussian system, for example,
one introduces effectively a new quantity
$\vec{d}:={\vec{D}}/{\varepsilon_0}$ and thereby finds in vacuum
$\vec{d}=\vec{E}$. However, $\vec{d}$ carries no longer the dimension
of an excitation and, accordingly, is only some superficial notational
construct without operational interpretation. The conventional vacuum
relation $\vec{D}=\vec{E}$, used in Gaussian units, defies a
reasonable dimensional analysis.

For ordinary linear and isotropic materials we can write
\begin{eqnarray}\label{const1}
  \vec{D}&=&\varepsilon_{\rm
    r}\varepsilon_0\vec{E}\,,\quad\vec{H}=\frac{1}{\mu_{\rm
      r}\mu_0}\vec{B}\,, \\ \nonumber
   [\varepsilon_0]& =& \frac{Y}{v}\,,\quad [\mu_0] = \frac{1}{Yv}\,, 
  \quad[\varepsilon_{\rm r}]=[\mu_{\rm r}]=1\,,
\end{eqnarray}
where $\vec{H}$ is the magnetic excitation and $\vec{D}$ the electric
excitation that are induced by the externally applied fields $\vec{E}$
and $\vec{B}$. We immediately recognize that the moduli characterizing
the material, namely the (relative) permittivity $\varepsilon_{\rm r}$
and the (relative) permeability $\mu_{\rm r}$ are both dimensionless,
quite independently of any system of units. This convenient
parameterization is achieved as soon as we introduce some standard and
constant $\varepsilon_0$ and $\mu_0$, with the dimensions of $Y/v$ and
$1/(Yv)$, respectively.

It turns out in a relativistic analysis of the Maxwell equations that
an admittance---or resistance as its reci\-procal---are
four--dimensional {\it scalar} quantities even in the Riemannian
spacetime of general relativity. This is why we concentrate on such a
quantity with a dimension of admittance. That a velocity is inherent
in electrodynamics is well-known and was already established by the
experiment of Weber-Kohlrausch before the advent of Maxwell's theory,
see \cite{LBrown}. This vacuum speed of light $c$ is a {\it
  special}-relativistic 4-dimen\-sional {\it scalar} as was uncovered
when Minkowski set up a manifestly Poincar\'e-covariant version of
electrodynamics (compare \cite{egoMinkowski}). Therefore, two
4-dimen\-sional scalar quantities, namely the {\it vacuum impedance}
$Y_0$ and the {\it vacuum speed of light} $c$ characterize the vacuum
from an electrodynamical point of view. In different system of units
they have different values, in SI, for instance, we have
$Y_0=\sqrt{\varepsilon_0/\mu_0}\approx 1/(377\,\Omega)$ and
$c=1/\sqrt{\varepsilon_0\mu_0}\approx 3\times 10^{8}\,{\rm m/s}$. But
the dimensionless permittivity/permea\-bility, $\varepsilon_{\rm r}$
and $\mu_{\rm r}$, keep their values in all systems of units.

\subsection{Four-dimensional excitation and field strength tensors}

In order to extend this characterization of materials by dimensional
quantities also to magnetoelectric moduli, we have first to turn to
the 4-dimensional representation of electrodynamics. Each of the two
sets $(\vec{D},\vec{H})$ and $(\vec{E},\vec{B})$ constitutes a
relativistic ``couple'' that is, we can define 4-dimensional 2nd rank
antisymmetric tensors according to ($\lambda,\nu=0,1,2,3$)
\begin{eqnarray}
\label{4tensor1}
\mathfrak{G}&=&(\mathfrak{G}^{\lambda\nu})=-(\mathfrak{G}^{\nu\lambda})=
\left(\begin{array}{rrrr}
 \hspace{-2pt}   0\hspace{5pt}& { D}^1 & { D}^2 & { D}^3\\ 
 \hspace{-2pt}    -{ D}^1 &0\hspace{5pt} & { H}_3 &
    -{ H}_2\\ \hspace{-2pt} -{ D}^2 & -{ H}_3 &0\hspace{5pt} & 
    { H}_1\\  \hspace{-2pt}-{ D}^3 & { H}_2 & -{ H}_1
    &0\hspace{5pt}
  \end{array}\right)\!,\\
\label{4tensor2} {F}&=&({F}_{\lambda\nu})=-({F}_{\nu\lambda})
=\;\left(\begin{array}{rrrr}\hspace{2pt}0\hspace{5pt} & -E_1 & -E_2 &
    -E_3\\ \hspace{2pt}E_1 &0\hspace{5pt} & B^3 & -B^2\\\hspace{2pt}
    E_2 & -B^3 &0\hspace{5pt} & B^1\\\hspace{2pt} E_3 & B^2 & -B^1
    &0\hspace{5pt}
  \end{array}\right)\!.
\end{eqnarray}

Even if one is not familiar with these expressions, we can see by a
dimensional analysis that each couple is really intrinsically
interrelated. We have $[\vec{H}]={I}/{\ell}={q}/
({t\ell})\,,\,[\vec{D}]={q}/{\ell^2}$, and
$[\vec{D}]=({1}/{v})\,[\vec{H}]$. Accordingly, $\vec{D}$ and $\vec{H}$
are both {\it charge-related} quantities distinguished only by the
dimension of velocity. This dimension of velocity mediates between
time and space components like for a coordinate where
$[x^0]=t,\,[x^1]=\ell=vt,...$, that is,
$[x^0]=(1/v)[x^1],\dots$ Analogously, we have
$[\mathfrak{G}^{01}]=[D^1]=(1/v)[H_1]=(1/v)[\mathfrak{G}^{23}]$ etc.


For the couple $(\vec{E},\vec{B})$ we have, with $\phi$ as
dimension of magnetic flux, $[\vec{E}]={U}/{\ell}={\phi}/({t\ell})$
and $[\vec{B}]={\phi}/{\ell^2}$, that is, $[\vec{E}]=v[\vec{B}]$. In
other words, $\vec{E}$ and $\vec{B}$ are {\it flux-related} (or, via
the Lorentz force, force-related) quantities. 

Apparently, the matrices in (\ref{4tensor1}) and (\ref{4tensor2}) can
be consistently set up from a dimensional point of view, including the
factors involving the velocity, provided the mentioned couples are
used. Thus, $[\mathfrak{G}]\sim$ charge, $[F]\sim$ flux, and
$[\mathfrak{G}]\times[F]\sim$ charge $\times$ flux = action; for a
detailed discussion see \cite{Birkbook}. We recognize thereby that the
scalar (density) with the dimension of an action,
\begin{equation}\label{Lagrangian}
  \frac 14\mathfrak{G}^{\lambda\nu}{F}_{\lambda\nu}=
  \frac 12\left(\vec{B}\cdot\vec{H}-\vec{D}
    \cdot\vec{E}\right)\,,
\end{equation}
must have a fundamental importance in electrodynamics: It represents
the Lagrangian of the electromagnetic field and is equally well a
4-dimensional general relativistic scalar.

In the literature these results are hardly recognized.
Landau-Lifshitz \cite{LL}, for example, like most of the engineers,
see Sihvola and Lindell \cite{SLAdP}, take the bastard pair
$(\vec{E},\vec{H})$ for their considerations. From a dimensional as
well as from a relativistic point of view, there is no legitimacy to
give birth to such a hybrid. All dimensions in relations get mixed up
by such an unholy pair.\footnote{There is one situation, however, in
  which the non-relativistic pair $(\vec{E},\vec{H})$ can be
  useful. Since in the Maxwell equations only $\vec{D}$ and $\vec{B}$
  enter as time derivatives, see ${\rm curl}\,\vec{H}=j+\dot{\vec{D}}$
  and ${\rm curl}\, \vec{E}+\dot{\vec{B}}=0$, the pair
  $(\vec{D},\vec{B})$ and by implication $(\vec{E},\vec{H})$ are
  useful for formulating the initial value problem in electrodynamics,
  as was pointed out by Perlick \cite{Perlick}. In this situation one
  has to leave the manifestly relativistic covariant formulation of
  electrodynamics and introduces necessarily a particular slicing of
  the 4-dimensional spacetime that destroys the manifest covariance of
  the theory.}

\subsection{Dimensionless magnetoelectric moduli}

According to (\ref{4tensor1}) and (\ref{4tensor2}), the general {\it
  local} and {\it linear} constitutive law is expected to be of the
general form $\mathfrak{G}\propto F$. The simplest case is that of an
isotropic medium. If then, in the magnetoelectric case, a magnetic
field $\vec{B}$ (or an electric field $\vec{E}$) is applied, the
additional response will be an electric excitation $\vec{D}$ (or a
magnetic excitation $\vec{H}$):
\begin{eqnarray}\label{const2}
  \vec{D}&=&\a_{\rm r}\a_0\, \vec{B}\,,\qquad
  \vec{H}=\b_{\rm r}\b_0\,\vec{E}\,;\\ \nonumber
  [\a_0]&=& [\b_0]=\frac{[\vec{D}]}{[\vec{B}]}
=\frac{[\vec{H}]}{[\vec{E}]}=Y\,,\quad 
[\a_{\rm r}]=[\b_{\rm r}]=1\,.
\end{eqnarray}

Since the dimensions of $\a_0$ and $\b_0$ are the same, we can put
them equally to the vacuum admittance: $\a_0=\b_0=Y_0$. If we allow an
electric and a magnetic field to be present at the same time, then
(\ref{const1}) combined with (\ref{const2}) for isotropic media leads
to the ansatz
\begin{eqnarray}\label{me1}
\vec{D}&=&Y_0\left( \varepsilon_{\rm
  r}\frac{1}{c}\,\vec{E}+\a_{\rm r}\,\vec{B}\right)\,,\\
\label{me2}
\vec{H}&=&Y_0\left(\frac{1}{\mu_{\rm
    r}}c\,\vec{B}+\b_{\rm r}\,\vec{E}\right)\,;\\
\label{me3}\nonumber
[Y_0]&=&{\rm admittance}\,,\quad [c]={\rm velocity},\\
\label{me4}\nonumber
 [\varepsilon_{\rm
    r}]&=&[\mu_{\rm
    r}]=[\a_{\rm r}]=[\b_{\rm r}]=1\,.
\end{eqnarray}
Note that $Y_0/c=\varepsilon_0$ and $Y_0c=1/\mu_0$.

Hence we see that all electric, magnetic, and magnetoelectic moduli,
independent of any system of units, can be described by the {\it
  dimensionless} quantities $\varepsilon_{\rm r},\,\mu_{\rm
  r},\,\a_{\rm r},\,\b_{\rm r}$. This has been shown, amongst others,
by Post \cite{Post1962}, O'Dell \cite{O'Dell1970}, Ascher
\cite{Ascher1970,Ascher1974a,Ascher1974b}, and Hehl et al.\
\cite{pseudoscalar}. Units such as those used in Ref.\cite{Fiebig2},
p.~R128, for example, namely $({\rm V}/{\rm cm})/{\rm O\!e}$ should be
abandoned. They complicate the comparison of different experiments.

If experimentalists collect magnetoelectric constants, then our
suggestion is {\em to tabulate them in their dimensionless relative
  form,} since these values are totally independent of the system of
units used and represent their genuine values.

\subsection{Local and linear magnetoelectric constitutive law}

Now that we have an elementary understanding of the dimensions of the
magnetoelectric moduli, we proceed to formulate, following Tamm
\cite{Tamm1925}, Post \cite{Post1962}, and our discussion in
\cite{pseudoscalar}, a 4-dimensional covariant law for a local,
linear, and anisotropic medium thereby generalizing the corresponding
isotropic law (\ref{me1}),(\ref{me2}),
\begin{equation}\label{linear}
  \mathfrak{G}^{\lambda\nu}=\frac
  12\,\chi^{\lambda\nu\sigma\kappa}F_{\sigma\kappa}=\frac
  12\,Y_0\,\xi^{\lambda\nu\sigma\kappa}F_{\sigma\kappa}\,,
\end{equation}
where the constitutive tensor $\chi$ of rank 4 and weight $+1$ is of
dimension $[\chi]=Y$; thus, $[\xi]=1$. Because of the antisymmetry of
$\mathfrak{G}^{\lambda\nu}$ and $F_{\sigma\kappa}$, we have
$\chi^{\lambda\nu\sigma\kappa}=-\chi^{\lambda\nu\kappa\sigma}=
-\chi^{\nu\lambda\sigma\kappa}$.  An antisymmetric pair of indices
corresponds, in four dimensions, to six independent components. Thus,
the constitutive tensor can be considered as a $6\times 6$ matrix with
36 independent components.

A $6\times 6$ matrix can be decomposed in its tracefree symmetric part
(20 independent components), its antisymmetric part (15 components),
and its trace (1 component). On the level of
$\chi^{\lambda\nu\sigma\kappa}$, this {\it decomposition} is reflected
in
\begin{eqnarray}\label{chidec}
  \chi^{\lambda\nu\sigma\kappa}&=&\,^{(1)}\chi^{\lambda\nu\sigma\kappa}+
  \,^{(2)}\chi^{\lambda\nu\sigma\kappa}+
  \,^{(3)}\chi^{\lambda\nu\sigma\kappa}\,.\\ \nonumber 36
  &=&\hspace{15pt} 20\hspace{15pt}\oplus \hspace{15pt}15\hspace{15pt}
  \oplus \hspace{25pt}1\,.
\end{eqnarray}

The third part, the {\it axion} part, is totally antisymmetric and as
such proportional to the totally antisymmetric Levi-Civita symbol, $
^{(3)}\chi^{\lambda\nu\sigma\kappa}:= \widetilde{\a}\,
\widetilde{\epsilon}^{\lambda\nu\sigma\kappa}$. We remind ourselves
that the Levi-Civita symbol $
\widetilde{\epsilon}^{\lambda\nu\sigma\kappa}=+ 1$ or $=-1$ depending
whether $\lambda\nu\sigma\kappa$ denotes an even or an odd permutation
of the numbers $0123$, respectively; it is zero otherwise, see
\cite{Sokol}. We denote pseudoscalars with a tilde.  

The second part, the {\it skewon} part, is defined according to $
^{(2)}\chi^{\mu\nu\lambda\rho}:=\frac 12(\chi^{\mu\nu\lambda\rho}-
\chi^{\lambda\rho\mu\nu})$.  If the constitutive equation can be
derived from a Lagrangian, which is the case as long as only
reversible processes are considered, then
$^{(2)}\chi^{\lambda\nu\sigma\kappa}=0$. The {\it principal} part
$^{(1)}\chi^{\lambda\nu\sigma\kappa}$ fulfills the symmetries $
^{(1)}\chi^{\lambda\nu\sigma\kappa}=
{}^{(1)}\chi^{\sigma\kappa\lambda\nu}$ and
$^{(1)}\chi^{[\lambda\nu\sigma\kappa]}=0$.  The constitutive relation
now reads
\begin{eqnarray}\label{constit7}
  \mathfrak{G}^{\lambda\nu}&=&
  \frac 12\left({}^{(1)}{\chi}^{\lambda\nu\sigma\kappa}+
    {}^{(2)}{\chi}^{\lambda\nu\sigma\kappa} +\widetilde{\a}\,
    \widetilde{\epsilon}^{\lambda\nu\sigma\kappa}\right)
  F_{\sigma\kappa}\nonumber\\
  &=&\frac 12 Y_0\left({}^{(1)}{\xi}^{\lambda\nu\sigma\kappa}+
    {}^{(2)}{\xi}^{\lambda\nu\sigma\kappa} +\widetilde{\a}_{\rm r}\,
    \widetilde{\epsilon}^{\lambda\nu\sigma\kappa}\right)
  F_{\sigma\kappa},
\end{eqnarray}
with $\widetilde{\a}_{\rm r}$ as dimensionless, $[\widetilde{\a}_{\rm
  r}]=1$.

In order to compare (\ref{constit7}) with experiments---the law
(\ref{me1}),(\ref{me2}) is a special case of it---we have to split
(\ref{constit7}) into time and space parts. As shown in
\cite{Postconstraint} in detail, we can parameterize the {\it
  principal} part by the 6 permittivities $\varepsilon^{ab}=
\varepsilon^{ba}$, the 6 permeabilities $\mu_{ab}=\mu_{ba}$, and the 8
magnetoelectric pieces $\g^a{}_b$ (its trace vanishes, $\g^c{}_c=0$,
summation over $c$!)  and the {\it skewon} part by the 3
permittivities $n_a$, the 3 permeabilities $m^a$, and the 9
magnetoelectric pieces $s_a{}^b$. Then (\ref{constit7}) can be
rewritten as ($a,b,c=1,2,3$)
\begin{eqnarray}\label{explicit'}\nonumber
  {D}^a&=&\left( \varepsilon^{ab}\hspace{4pt} - \,
    \epsilon^{abc}\,n_c \right)E_b\,\\ &&\hspace{25pt} 
  +\left(\hspace{6pt} \gamma^a{}_b +
    s_b{}^a - \delta_b^a s_c{}^c\right) {B}^b +
  \widetilde{\alpha}\,B^a \,, \\ {H}_a&=&\left( \mu_{ab}^{-1}\nonumber
    - \hat{\epsilon}_{abc}m^c \right) {B}^b\\ &&\hspace{25pt}  
+\left(- \gamma^b{}_a +
    s_a{}^b - \delta_a^b s_c{}^c\right)E_b -
  \widetilde{\alpha}\,E_a\,.\label{explicit''}
\end{eqnarray}
Here $\epsilon^{abc}= \hat{\epsilon}_{abc}=\pm 1,0$ are the
3-dimensional Levi-Civita symbols. As can be seen from our derivation,
$\widetilde{\alpha}=\widetilde{\alpha}_{\rm r}Y_0$ is a 4-dimensional
pseudo- (or axial) scalar, whereas $s_c{}^c$ is only a 3-dimensional
scalar. The cross-term $\gamma^a{}_b$ is related to the Fresnel-Fizeau
effects.  The skewon contributions $m^c,n_c$ are responsible for
electric and magnetic Faraday effects, respectively, whereas the
skewon terms $s_a{}^b$ describe optical activity. Equivalent
constitutive relations were formulated, amongst others, by Serdyukov
et al.\ \cite{Serdyukov}, by Lindell \cite{Ismobook}, and by Spaldin
et al.\ \cite{Nicola}. Magnetoelectric effects in spiral magnets and
in ferromagnets were recently studied by Mostovoy \cite{Mostovoy} and
Dzyaloshinskii \cite{Dz2008}, respectively.

From (\ref{explicit'}),(\ref{explicit''}), for
$\varepsilon^{ab}=\varepsilon g^{ab}$, where $g^{ab}$ are the
components of the 3-dimensional (contravariant) metric,
$(\mu^{-a})_{ab}$ $=(\mu^{-1})g_{ab}$, and all other magnetoelectric
moduli vanishing, we recover the special case
\begin{eqnarray}\label{iso1}
D^a&=&\hspace{8pt}\varepsilon E^a\,\hspace{8pt} +\widetilde{\a}\, B^a\,,\\
H_a&=& (\mu^{-1})B_a-\widetilde{\a} E_a\,,\label{iso2}
\end{eqnarray}
with $E^a=g^{ab}E_b$ and $B_a=g_{ab}B^b$. This is the isotropic case
of (\ref{me1}),(\ref{me2}); but additionally, we recognize that in
(\ref{me2}) we have to require \begin{equation}\label{xxx}\b_{\rm
    r}=-\a_{\rm r}\end{equation} for consistency.

Incidentally, if necessary, the linear law (\ref{linear}) can be
amended with a {\it quadratic} term according to the scheme
\begin{equation}\label{quadratic}  
  \mathfrak{G}^{\lambda\nu}=\frac
  12\,\chi^{\lambda\nu\sigma\kappa}F_{\sigma\kappa}+
  \frac 14
  \eta^{\lambda\nu\sigma\kappa\mu\tau} F_{\sigma\kappa} F_{\mu\tau}\,,
\end{equation}
with an additional constitutive tensor $\eta$ of 126 independent
components. In the reversible case they reduce to 56 independent
components.

\section{The relativistic pseudoscalar $\widetilde{\a}$}

\subsection{Violation of the Post constraint}

According to some hand-waving arguments of Post, see \cite{Post1962},
p.~129, the pseudoscalar $\widetilde{\a}$ in (\ref{explicit'}),
(\ref{explicit''}) (and even its first derivative) should vanish for
the vacuum as well as for all media.  Lakhtakia, see
\cite{Akhlesh2004}, dubbed the relation $\widetilde{\a}=0$ the ``Post
constraint'' and pushed it as a presumed consequence of Maxwell's
equations quite vigorously. Lakhta\-kia's arguments have never been
convincing to a determined minority, see, for instance, Sihvola and
Tretyakov \cite{Ari1995,SihTre}. More recently Lakhtakia
\cite{Akhlesh2009} admitted that he was driven to the Post constraint
by the following two prejudices \cite{Akhlesh2009}:

``(i) The idea of a nonreciprocal but isotropic medium appears
oxymoronic [contradictory].''

``(ii) The idea of a constitutive parameter that vanishes from the
macroscopic Maxwell equations for a linear medium but leaves behind
its spatiotemporal derivatives appears bizarre.''

To (i): This is the case of (\ref{iso1}),(\ref{iso2}). If a material
is ``charged'' with the 4-dimensional {\it pseudo\/}scalar
$\widetilde{\a}$---admit\-tedly a case that is rare in nature---then an
orientation dependence of $\widetilde{\a}$ is implied; why should this
contradict the 3-dimensional isotropy of the medium?

To (ii): If $\widetilde{\a}$ is spacetime dependent, $\widetilde{\a}=
\widetilde{\a}(x)$, then it emerges in the field equations of
electrodynamics, as had already been shown by Wilczek \cite{Frank1987}
in his axion-electrodynamics. In the constant case, the
$\widetilde{\a}$ jumps at the interface between the medium and vacuum
and yields a contribution to the corresponding boundary conditions,
see \cite{measuringAxion}. Moreover, at least since the very exact
measurements of Wiegelmann et al.\ \cite{Wiegelmannetal} no doubt was
possible at the violation of the Post constraint in the case of the
antiferromagnet Cr$_2$O$_3$.

\subsection{Dzyaloshinskii's theory for the magnetoelectric effect of
  the antiferromagnet Cr$_2$O$_3$}\label{sec3.2}

Landau and Lifshitz (1956, see the first Russian edition of
\cite{LL}) predicted the magnetoelectric effect as a phenomenon that
``results from a linear relation between the magnetic and the electric
fields in a substance, which would cause, for example, a magnetization
proportional to the electric field...'' It ``can occur for certain
magnetic crystal symmetry classes.'' Dzyaloshinskii \cite{Dz1}, on the
basis of neutron scattering data and susceptibility measurements,
found out that the antiferromagnetic chromium sesquioxide Cr$_2$O$_3$
has the desired magnetic symmetry class. Starting from a thermodynamic
potential quadratic and bilinear in $\mathbf{E}$ and $\mathbf{H}$, he
developed a theory for the magnetoelectic constitutive relations for
Cr$_2$O$_3$. Written as quantity equations, they are valid in an
arbitrary system of units and read
\begin{eqnarray}\label{DH1}
  D_x &=& \varepsilon_{\bot} \frac{Y_0}{c} E_x
  +\alpha_{\bot}\frac 1c H_x\,,\\ \label{DH2} D_y &=&
  \varepsilon_{\bot} \frac{Y_0}{c} E_y +\alpha_{\bot}
  \frac 1c H_y\,,\\ \label{DH3} D_z &=&
  \varepsilon_{||}\,\frac{Y_0}{c} E_z +\alpha_{||}\,\frac 1c H_z\,
\end{eqnarray}
and
\begin{eqnarray}\label{BE1}
  B_x &=& \mu_{\bot}\frac{1}{Y_0 c} H_x +\alpha_{\bot}\frac{1}{c}
  E_x\,,\\ B_y &=& \mu_{\bot}\frac{1}{Y_0 c} H_y
  +\alpha_{\bot}\frac{1}{c} E_y\,, \label{BE2}
\\ \label{BE3} B_z &=&
  \mu_{||}\,\frac{1}{Y_0 c} H_z +\alpha_{||}\,\frac 1c E_z\,.
\end{eqnarray}
The $z$-axis is parallel to the optical axis of Cr$_2$O$_3$.  Here we
have (relative) permittivities parallel and perpendicular to the
z-axis of the crystal, namely $\varepsilon_{||},\varepsilon_{\bot}$,
analogous (relative) permeabilities $ \mu_{||}, \mu_{\bot}$ and
magnetoelectric moduli $\alpha_{||},\alpha_{\bot}$.

As we can see from (\ref{explicit'}) and (\ref{explicit''}), we have
to get $(\mathbf{D},\mathbf{H})$ on the left hand side and
$(\mathbf{E},\mathbf{B})$ on the right hand side in order to end up
with a constitutive law that is written in a relativistically
covariant form. We find,
\begin{eqnarray}\label{Dcon1}
  D_x &=&Y_0\left[ \left(\varepsilon_{\bot}-\frac{\a_\bot^2}{\mu_\bot}\right)
 \frac 1c E_x +\frac{\alpha_{\bot}}{\mu_\bot} B_x\right]\,,\\ D_y
  &=&Y_0\left[\left( \varepsilon_{\bot}-\frac{\a_\bot^2}
    {\mu_\bot}\right) \frac 1c E_y +\frac{\alpha_{\bot}}
  {\mu_\bot} B_y\right]\,,\\ D_z &=&Y_0\left[\left(
   \varepsilon_{||}-\frac{\a_{||}^2}{\mu_{||}}\right)\,
   \frac 1c E_z +\frac{\alpha_{||}}{\mu_{||}}\,
  B_z\right]\,\label{Dcon3}
\end{eqnarray}
and 
\begin{eqnarray}\label{HB1} 
  H_x &=&Y_0\left[ \frac{1}{\mu_\bot}c B_x
  -\frac{\alpha_{\bot}}{\mu_\bot} E_x\right]\,,\\ \label{HB2} H_y &=&Y_0\left[
  \frac{1}{\mu_\bot}c B_y -\frac{\alpha_{\bot}}{\mu_\bot}
  E_y\right]\,,\\ 
  \label{HB3} H_z &=&Y_0\left[ \frac{1}{\mu_{||}}\,c B_z
  -\frac{\alpha_{||}}{\mu_{||}}\, E_z\right]\,.
\end{eqnarray}

We can compare these equations with (\ref{explicit'}) and
(\ref{explicit''}). Since Dzyaloshinskii assumed reversibility, the
skewon piece with its 15 independent components vanishes
identically. Hence  (\ref{explicit'}) and
(\ref{explicit''}) reduce to 
\begin{eqnarray}\label{explicit3}
  {D}^a\!&=\!& {\varepsilon^{{ab}}}\,E_b + {\gamma^a{}_b}\, {B}^b +
  {\widetilde{\a}}\,B^a \,,\\ {H}_a\!  &=\!  & { \mu_{ab}^{-1}} {B}^b - {
    \gamma^b{}_a}E_b - {\widetilde{\a}}\,E_a\,,\label{explicit4}
\end{eqnarray}
with 21 independent moduli. The 4-dimensional pseudo\-scalar
$\widetilde{\a}$ (alternatively called the axion parameter) represents
1 component. We compare (\ref{explicit3}) and (\ref{explicit4}) with
(\ref{Dcon1}) to (\ref{HB3}) and note that Dzyaloshinskii used
Cartesian coordinates such that $B^x=B_x$ etc. Then we find the
permittivity
\begin{equation}\label{perm1}
  \varepsilon^{ab}= \frac{Y_0}{c}\begin{pmatrix}
    \varepsilon_\bot-\frac{\a_\bot^2}{\mu_\bot}&0&0\cr 0&
    \varepsilon_\bot-\frac{\a_\bot^2}{\mu_\bot}&0\cr 0&0&
    \varepsilon_{||}-\frac{\a_{||}^2}{\mu_{||}} \end{pmatrix}\,,
\end{equation}
the impermeability
\begin{equation}\label{perm2}
  \mu^{-1}_{ab}= Y_0 c\begin{pmatrix} \mu^{-1}_\bot&0&0\cr 0&
  \mu^{-1}_\bot&0\cr 0&0& \mu^{-1}_{||}\end{pmatrix}\,,
\end{equation}
the magnetoelectric $\gamma$ matrix
\begin{equation}\label{gamma}
  \gamma^a{}_b= \frac 13\left(\frac{\a_\bot}{\mu_\bot}- 
\frac{\a_{||}}{\mu_{||}} \right)\sq2 
  \begin{pmatrix}1&\hspace{4pt}0&\hspace{4pt}0\cr
    0&\hspace{4pt}1&\hspace{4pt}0\cr
    0&\hspace{4pt}0&\hspace{-2pt}
-2 \end{pmatrix}\,,
\end{equation}
and the pseudoscalar (or axion) piece
\begin{equation}\label{axion}
    \widetilde{\a}= \frac 13\left(2\,\frac{\a_\bot}{\mu_\bot}+ 
      \frac{\a_{||}}{\mu_{||}} \right)\sq2 \,.
\end{equation}

Conventionally, in the magnetoelectric literature the $\g$-matrix
and $\widetilde{\a}$ are collected in the ``relativistic'' matrix
\begin{equation}\label{alpha}
^{\rm rel}\a^a{}_b:=\g^a{}_b+\widetilde{\a}\,\delta^a_b
=\sq2\begin{pmatrix}
\frac{\a_\bot}{\mu_\bot}&0&0\\
0&\frac{\a_\bot}{\mu_\bot}&0\\
0&0&\frac{\a_{||}}{\mu_{||}}\end{pmatrix}\,.
\end{equation}
We called it relativistic, since it occurs in the context of the
relativistic $(\mathbf{E},\mathbf{B})$ system.  Later, the general
form of these matrices for the crystal structure of Cr$_2$O$_3$ was
confirmed by O'Dell \cite{O'Dell1970}, amongst others. Note that
$\ve_{||},\ve_\bot,\mu_{||},$ $\mu_\bot, \a_{||}$, and $\a_\bot$,
according to their definitions (\ref{DH1}) to (\ref{BE3}), are
measured in an external $\mathbf{E}$ and/or an external $\mathbf{H}$
field.

\subsection{Experiments of Astrov, Rado \& Folen, and Wiegelmann et
  al.}\label{Sec.3.3}

The magnetoelectric effect for Cr$_2$O$_3$ was first found
experimentally by Astrov \cite{Astrov1961} for the {\it electrically}
induced magnetoelectric effect (called ME$_{\rm E}$ in future) and by
Rado \& Folen \cite{RadoFolen1961} for the {\it magnetically} induced
magnetoelectric effect (ME$_{\rm H}$). In both investigations single
crystals of Cr$_2$O$_3$ were used. In the ME$_{\rm E}$ experiments
\cite{Astrov1961,RadoFolen1961}, Eqs.(\ref{BE1}) to (\ref{BE3}) were
verified ($H$ switched off) and in the ME$_{\rm H}$ experiments
\cite{RadoFolen1961} Eqs.(\ref{DH1}) to (\ref{DH3}) ($E$ switched
off). In particular, Rado \& Folen made both type of experiments and
found that the magnetoelectric moduli $\a_\bot$ and $\a_{||}$ for
ME$_{\rm E}$ experiments coincide with those of the ME$_{\rm H}$
experiments. This proves the vanishing of the skewon part
$^{(2)\!}\chi^{\lambda\nu\sigma\kappa}$ of the constitutive tensor
$\chi^{\lambda\nu\sigma\kappa}$ for Cr$_2$O$_3$.

Accordingly, these experiments confirmed Dzyaloshinskii's theory for
Cr$_2$O$_3$ below the spin-flop phase. Further experiments were then
done mainly for the ME$_{\rm H}$ case. Particular accurate
measurements are those of Wiegelmann et al.\ \cite{Wiegelmannetal},
see also \cite{WiegelmannDr,Wiegelmann2}.  They took magnetic fields
$B$ as high as 20 {tesla} and measured from liquid Helium up to room
temperature. Wiegelmann et al.\ took a quasi-static magnetic field and
thereby proved explicitly that measurements with magnetic fields of
some kilo hertz can be extrapolated to static measurements. The sign
of $\a_\bot(T)$ relative to $\a_{||}(T)$ was left open in
\cite{Wiegelmannetal}. Hence we took that from Astrov
\cite{Astrov1961}.

We plotted the values of $\a_\bot(T)$ and $\a_{||}(T)$ from the
references quoted. We have shown in \cite{pseudoscalar} that the
permeabilities $\mu_\bot\approx \mu_{||}\approx 1$. Then we can
determine $\wa$ by means of (\ref{axion}). The values we found
\cite{pseudoscalar} are displayed in Fig.~1. Up to about 163 K, the
pseudoscalar $\wa$ is negative, for higher temperatures positive until
it vanishes at the N\'eel temperature of about 308 K. Its maximal
value we find at 285 K:
\begin{eqnarray}\label{result1}
  \widetilde{\a}_{{\rm max}}\;\mbox{(at 285 $K$)}&\approx& 3.10\times 
  10^{-4}\;\sq2\,.
\end{eqnarray}
Consequently, we have definitely a nonvanishing pseudo\-scalar
$\wa=\wa_{\rm r}Y_0$ with a maximal $\wa_{\rm r}$ of the order of
$10^{-4}$. It is a small effect, but it does exist.

\begin{figure}\label{fig9}
\includegraphics
[width=9cm,height=6.5cm]{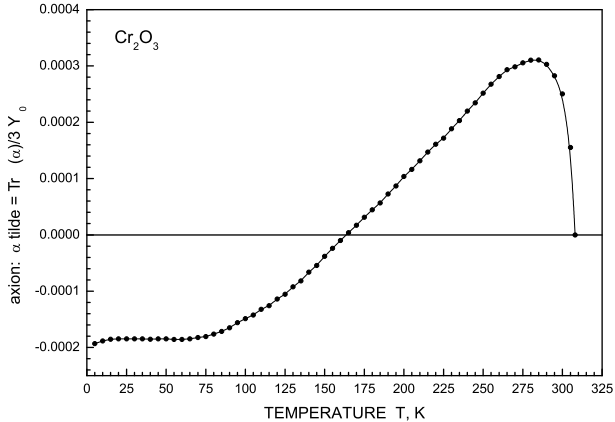}
\caption{The pseudoscalar (or axion) piece $\widetilde{\a}$ of
  Eq.~(\ref{axion}) of Cr$_2$O$_3$ in units of the vacuum admittance
  $Y_0$ as a function of the temperature $T$ in kelvin; in SI,
  $Y_0\approx 1/(377\,{\rm ohm})$; for details see
  \cite{pseudoscalar}.}
\end{figure}

\subsection{New predictions}

\subsubsection{Search for the first {\em cubic} magnetoelectric
  crystal}

As we saw in Sec.~\ref{Sec.3.3}, the parallel and perpendicular
permeabilities for Cr$_2$O$_3$ are approximately equal,
$\mu_{||}\approx\mu_{\bot}=\mu$. Then the magnetoelectric $\g$ matrix
of (\ref{gamma}) becomes
\begin{equation}\label{gamma'}
  \gamma^a{}_b\approx \frac{Y_0}{3}\frac{\a_\bot-\a_{||}}{\mu_{\rm r}} 
  \begin{pmatrix}1&\hspace{4pt}0&\hspace{4pt}0\cr
    0&\hspace{4pt}1&\hspace{4pt}0\cr
    0&\hspace{4pt}0&\hspace{-2pt}
-2 \end{pmatrix}\,.
\end{equation}
The question is now: Can we find a substance in which
\begin{equation}\label{subst}
{\a_\bot}={\a_{||}}\,,
\end{equation}
that is, in which the matrix $\g^a{}_b$ vanishes {\em for all
  temperatures?} This would be interesting to know since then one
would have a substance, like in (\ref{iso1}) and (\ref{iso2}), in
which the {\em only} magnetoelectric piece would be the pseudoscalar
(or axion) piece $\wa$.

We can take from Rivera \cite{Riveraunits}, Fig.2, box 16, that three
equal entries in the diagonal of the matrix (\ref{gamma'}) are only
possible in the following five antiferromagnetic {\it cubic} point groups:
\begin{equation}\label{five}
{\rm 23,\;\; m'\overline{3}{}',\;\; 432,\;\; \overline{4}{}'3m',\;\;
m'\overline{3}{}'m'\,.}
\end{equation}
Schmid \cite{B}, p.23, has explained that the {\it Cr-Br- and
  Cr-I-boracites\/} are antiferromagnetic and cubic below $17\,$K and
$54\,$K, respectively. The Cu-I boracite stays also cubic down to very
low temperatures \cite{[A]} and is expected to behave analogously to
Cr-I boracite, since both Cr$^{++}$ and Cu$^{++}$ are Jahn-Teller
ions. However, the N\'eel temperature of Cu-I boracite is not known so
far. Due to the cubic paramagnetic prototype point group ${\rm
  \overline{4}3m1'}$, only two antiferromagnetic point groups are
allowed:
\begin{equation}\label{two}
{\rm \overline{4}{}'3m'\quad\mbox{and} \quad \overline{4}3m\,.}
\end{equation}
Whereas the point group ${\rm \overline{4}{}'3m'}$ allows the linear
magnetoelectric (ME) effect (invariant $\overline{\a}_{ab} E_a H_b$)
and both higher order ME effects (invariants $\overline{\g}_{abc}H_a
E_b E_c$ and $\overline{\b}_{abc}E_a H_b H_c$), the point group
${\rm\overline{4}3m}$ allows only the invariant
$\overline{\b}_{abc}E_a H_b H_c$ (see \cite{[B]}, Table 2). It could
be that {\it one or more of these three boracites have} ${\rm
  \overline{4}{}'3m'}$ {\it symmetry.} 

In order to check this, ME measurements on single crystals would be
desirable. However, the linear ME effect can in principle also be
detected on compacted, magnetoelectrically annealed, polycrystalline
powders or ceramics \cite{[C],[D],[E]}, both by ME$_{\rm H}$ and
ME$_{\rm E}$ methods. The $\overline{\b}_{abc}E_a H_b H_c$-effect
cancels out statistically in powders and ceramics because the sign of
the coefficient is rigorously linked to the orientation of the
non-centrosymmetric, nuclear absolute structure (note, only in poled
ferroelectric ceramics this statistics is broken and the effect
becomes measurable in principle, when using the ME$_{\rm H}$
effect). Because the $\overline{\g}_{abc}H_a E_b E_c$-effect cannot be
measured by a ME$_{\rm H}$ quasistatic method, it would not disturb
when one measures the linear effect of a phase of the point group
${\rm \overline{4}{}'3m'}$. The $\overline{\g}_{abc}H_a E_b
E_c$-effect can be measured by a ME$_{\rm E}$ method or by a ME$_{\rm
  H}$ technique in presence of a static electric field, as first shown
by O'Dell \cite{[A']}. As a consequence, even ME measurements on
compacted powders or on ceramics would allow to distinguish between
the two point groups, with ${\rm\overline{4}3m}$ showing no ME effect
at all and with ${\rm \overline{4}{}'3m'}$ only the linear effect when
a ME$_{\rm H}$ quasistatic method is used. We would like to suggest to
grow single crystals \cite{[F],[G]} and to perform ME$_{\rm H}$
measurements \cite{[B']} both on single crystals and polycrystalline
samples. A polycrystalline probe of a cubic magnetoelectric crystal is
obviously completely isotropic for the linear magnetoelectric
effect. Hence with a view on applications, the production of
polycrystalline material could be simpler than growing single
crystals.

\subsubsection{ External magnetic {\em octupole\/} field of a cubic
  magnetoelectric crystal}

Dzyaloshinskii \cite{Dzpriv} hypothesized that there may exist an
external magnetic field for free cubic (or axion type) magnetoelectric
media.  Normally, the external magnetic field outside an
antiferromagnet decays exponentially. Much earlier, Dzyaloshinskii
\cite{Dzyal} predicted the existence of an external magnetic and
electric field for magnetoelectric media with high symmetry, and
specifically for the antiferromagnetic Cr$_2$O$_3$. Such a field
decreases according to a power law from the surface of a body, in
contrast to the exponential decay typical for substances with lower
symmetries. For Cr$_2$O$_3$, this external magnetic field behaves as
the field of a magnetic {\it quadrupole,} which was subsequently
confirmed experimentally by Astrov et al.\ \cite{Astr1,Astr2}.

For the magnetoelectric crystal with a cubic symmetry, which is
characterized by the purely axion piece in the linear constitutive
relation, we expect the similar effect \cite{Dzpriv}. Taking, for
definiteness, a ball made of such a material (recall that Astrov et
al.\ in the 1960s used little balls made out of the antiferromagnetic
Cr$_2$O$_3$), one can perform a usual expansion of the electric and
magnetic fields into spherical harmonics. From a preliminary
qualitative analysis of the surface charges and currents induced, one
can expect an external magnetic field of the {\em octupole\/}
structure, namely,
\begin{equation}
  B \sim -\,\widetilde{\alpha}\,\nabla\left({\frac {{\cal Y}_{4m}}{r^5}}\right),
\end{equation}
which is directly proportional to the axion piece $\widetilde{\alpha}$
(here, as usual, the ${\cal Y}_{lm}(\theta,\varphi)$ denote the
spherical harmonics). This preliminary result is based on the fact
that ${\cal Y}_{4m}$ is invariant under the cubic symmetry group of
the prospective cubic axion material.

\subsection{Relations of $\wa$ to gyrator, axion, and perfect
  electromagnetic conductor (PEMC)}

The constitutive law for the pseudoscalar piece, namely (\ref{const2})
together with (\ref{xxx}), reads
\begin{eqnarray}\label{iso1/2}
D^a&=& \hspace{7pt}\widetilde{\a}\, B^a,\nonumber\\
H_a&=& -\widetilde{\a} E_a\,,\label{iso2/2}
\end{eqnarray}
see also (\ref{iso1}), (\ref{iso2}). This structure is not
unprecedented, see the discussion in \cite{Postconstraint}. In
electrical engineering, in the theory linear networks, more
specifically in the theory of two ports (or four poles), Tellegen
\cite{Tellegen1948,Tellegen1956/7} invented the {\it gyrator} the
defining relations of which are
\begin{eqnarray}\label{gyrator}
&&v_1=-s\,i_2\,,\nonumber\\
&&v_2=\hspace{7pt}s\,i_1\,,
\end{eqnarray}
where $v$ are voltages and $i$ currents of the ports 1 and 2,
respectively, see also \cite{O'Dell1970}. Let us quote from Tellegen
\cite{Tellegen1956/7}, p.189: ``The ideal gyrator has the property of
`gyrating' a current into a voltage, and vice versa.  The coefficient
$s$, which has the dimension of a resistance, we call the gyration
resistance; $1/s$ we call the gyration conductance.'' The gyrator is a
nonreciprocal network element that is discussed in the electrical
engineering literature \cite{Feldtkeller,Kuepfmueller}, for more
recent developments see, e.g., \cite{RusserGyro} and \cite{Zhai}.

For dimensional reasons, the electromagnetic excitations $(D^a,H_a)$ are
related to the currents and the field strengths $(E_a,B^a)$ to the
voltages, that is,
\begin{eqnarray}\label{gyrator1}
&&E_a=-s\,H_a\,,\nonumber\\
&&B^a=\hspace{7pt}s\,D^a.
\end{eqnarray}
If we put $s=1/\widetilde{\a}$, then (\ref{gyrator1}) and
(\ref{iso1/2}) coincide. Similar as the gyrator rotates currents into
voltages, the axion piece `rotates' the excitations, modulo a
resistance, into the field strengths.

The next ``isomorphic'' structure to discuss is {\it axion
  electrodynamics,} see Ni \cite{Ni}, Wilczek \cite{Frank1987}, and,
for more recent work, Itin \cite{Itin2004,Itin2008,Itin2007}. We add
to the usual Maxwell-Lorentz law for vacuum electrodynamics
$\mathfrak{G}^{\lambda\nu}=Y_0\,\sqrt{-g}F^{\lambda\nu}$ an axion
piece patterned after the last term in (\ref{constit7}), then we have
the constitutive law for axion electrodynamics,
\begin{equation}\label{axel}
  \mathfrak{G}^{\lambda\nu}=Y_0\,\sqrt{-g}F^{\lambda\nu}+\frac
  12\,\widetilde{\a}\, \widetilde{\epsilon}^
  {\lambda\nu\sigma\kappa}F_{\sigma\kappa}\,.
\end{equation}

In order to get a feeling for the Lagrangian of axion electrodynamics,
we substitute (\ref{iso1}), (\ref{iso2}), and  (\ref{xxx}) into the
right-hand-side of (\ref{Lagrangian}); after some algebra we find
\begin{eqnarray}\label{Lagrangian1}
  \frac 14\mathfrak{G}^{\lambda\nu}{F}_{\lambda\nu}&=&
  \frac 12\left(\vec{B}\cdot\vec{H}-\vec{D}
    \cdot\vec{E}\right)\nonumber\\
  &=&\underbrace{ \frac 12\frac{1}{\mu_{\rm r}\mu_0}\vec{B}^2
-\frac 12\varepsilon_{\rm
      r}\varepsilon_0\vec{E}^2}_{\text{Maxwell-Lorentz part}}
\underbrace{-\wa_{\rm r}Y_0\,\vec{E}\cdot\vec{B}}_{\text{axion part}}\,.
\end{eqnarray}

As we saw, in Cr$_2$O$_3$ we had $\widetilde{\a}_{\rm r}\approx
10^{-4}$. It is everybody's guess what it could be in elementary
particle theory. There, in spite of extented searches via the coupling
of photons to the hypothetical pseudoscalar axion field, amongst other
methods, nothing has been found so far, see the review of Carosi et
al.\ \cite{Carosi}.

These isomorphisms even found a further field of applications: In
2005, Lindell \& Sihvola \cite{LindSihv2004a,LindSihv2004b}, see also
\cite{Ismobook}, introduced the new concept of a {\it perfect
  electromagnetic conductor} (PEMC). It also obeys the constitutive
law (\ref{iso1/2}). Some applications were studied by Jancewicz
\cite{Bernard} and by Illahi and Naqvi \cite{Illahi}, see also the
references given therein. The PEMC is a generalization of the perfect
electric and the perfect magnetic conductor. In this sense, it is the
`ideal' electromagnetic conductor that can be hopefully built by means
of a suitable {\it metamaterial,} see Sihvola \cite{metaAri}. The
pseudoscalar $\widetilde{\a}$ is called Tellegen parameter by Lindell
et al., see \cite{Lindell1994}, p.13 (for a more general view, see
\cite{SihvolaLindell1995}); artificial Tellegen material has been
produced and positively tested by Tretyakov et al.\ \cite{Tretyakov},
amongst others.

Accordingly, (i) the gyrator, (ii) the axion field of axion
electrodynamics, (iii) the pseudoscalar $\wa$ of Cr$_2$O$_3$, and (iv)
the PEMC are isomorphic structures. Wilczek (private communication)
remarked to these isomorphisms between these four systems: ``It's a
nice demonstration of the unity of physics.'' Already in his paper of
1987 \cite{Frank1987} he argued that ``...it is...not beyond the realm
of possibility that fields whose properties partially mimic those of
axion fields can be realized in condensed-matter systems.'' ...This
is, indeed, the case: The pseudoscalar $\wa$ of Cr$_2$O$_3$ is the
structure Wilczek looked for.

\begin{acknowledgement}
  We would like to thank the organizers, in particular Nicola Spalding
  (Santa Barbara), for giving us the possibility to present a seminar
  at their workshop. We are grateful to Nicola Spalding for pointing
  out to us reference \cite{WFBrown}.
\end{acknowledgement}

\end{document}